\documentclass[a4paper]{jpconf}
\usepackage{graphicx}
\usepackage{natbib}
\begin{document}
\title{Quantifying inhomogeneities in the HI distributions of simulated galaxies}

\author{Hind Al Noori, Andrea V. Macci{\`o}, Aaron A. Dutton, and Keri L. Dixon}

\address{New York University Abu Dhabi, Abu Dhabi, UAE}

\ead{hindalnoori@nyu.edu}

\begin{abstract}
The NIHAO cosmological simulations form a collection of a hundred high-resolution galaxies. We used these simulations to test the impact of stellar feedback on the morphology of the HI distribution in galaxies. We ran a subsample of twenty of the galaxies with different parameterizations of stellar feedback, looking for differences in the HI spatial distribution and morphology. We found that different feedback models do leave a signature in HI, and can potentially be compared with current and future observations. These findings can help inform future modeling efforts in the parameterization of stellar feedback in cosmological simulations of galaxy formation and evolution.
\end{abstract}

\section{Introduction}
We are currently in the ``golden age" of cosmology, with increasingly accurate measurements and constraints being placed on almost all cosmological parameters, painting a much richer conception of the Universe than was possible only a few decades ago. Despite that, a full understanding of the processes behind galaxy formation continues to evade us and has proven to be a difficult problem in modern astrophysics. A very common and powerful tool used to study galaxy formation is via cosmological hydrodynamical simulations that allow us to examine the interplay of cosmological parameters such as dark matter with the physics of standard baryonic matter. Fortunately, the abundance of well-constrained observational relations, such as the $M_{\rm star}-M_{\rm halo}$ relation, the $M$-$\sigma$ relation, and the Tully-Fisher relation, allow for easy comparison with relations derived by analyzing simulated galaxies, and can serve as sanity checks of the physical conditions in the simulations.

Many cosmological simulations of galaxy formation have been performed in recent years, each inhabiting its own niche in the parameter space of possible physics. Historically, cosmological simulations typically formed an excess of stars (in comparison to observation), often with a large concentration of stars at the center of a galaxy \citep[this problem is called the overcooling problem, ][]{stinson13}. Stellar feedback, in the form of supernova feedback and the winds of massive stars, was later discovered as a crucial ingredient in the physics of galaxy formation and has been shown to greatly alleviate the problem of the over-production of stars \citep{stinson13}. In this study we used the NIHAO (Numerical Investigation of a Hundred Astrophysical Objects) hydrodynamical zoom-in simulations, which are a unique collection of a hundred high-resolution galaxies spanning a range of galaxy types, with dark matter halo masses ranging from $\sim 10^9 M_{\odot}$ to $\sim 10^{12} M_{\odot}$ \citep{wang15}. The NIHAO simulations have been shown to agree with a number of observational relations, such as the inefficiency in star formation \citep{wang15} and the Tully-Fisher relation \citep{dutton17}.

\begin{figure}[h]
\begin{minipage}{0.33\textwidth}
\includegraphics[width=0.98\textwidth]{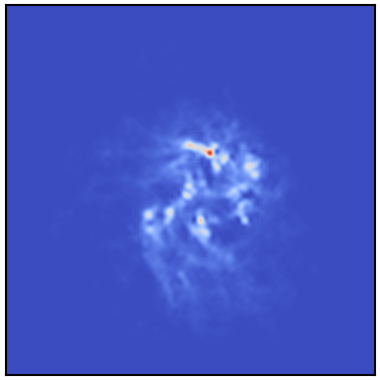}
\end{minipage}\hspace{0.1pc}%
\begin{minipage}{0.33\textwidth}
\includegraphics[width=0.98\textwidth]{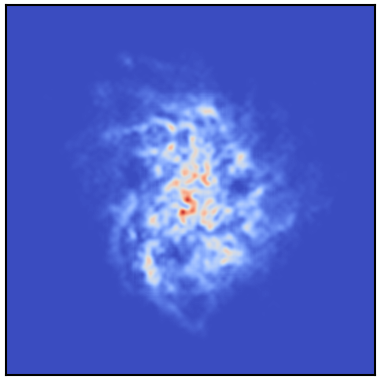}
\end{minipage}\hspace{0.1pc}%
\begin{minipage}{0.33\textwidth}
\includegraphics[width=0.98\textwidth]{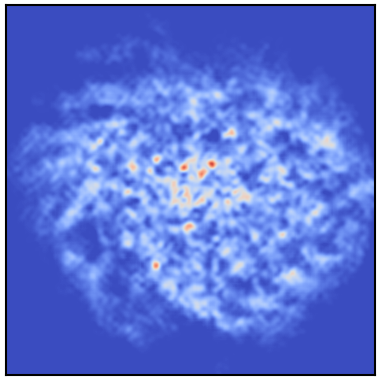}
\end{minipage} 
\caption{\label{galaxy} HI gas from a simulated galaxy at different star formation density thresholds $n$ = 10~cm$^{-3}$ (left), $n$ = 1.0~cm${-3}$ (center), and $n$ = 0.1~cm$^{-3}$ (right).}
\end{figure}

Star formation in simulations is ``induced" when a certain threshold density is reached in the gas. In the NIHAO simulations, a high particle density threshold of $n$ = 10~cm$^{-3}$ is used as a parametrization of star formation, compared to the high-volume cosmological simulations ILLUSTRIS and EAGLE that employ $n$ = 0.1~cm$^{-3}$ \citep{vogels14, schaye15}. Real stars form at densities of above $n = 10^4$ ~cm$^{-3}$ in molecular clouds, which is not feasible to replicate due to prohibitive computational costs in state-of-the-art simulations today. This calls for in-depth studies of the effects of different thresholds on galaxy formation and for attempts at relating them to observational data. In this study, for a subsample of 20 of the 100 NIHAO galaxies, we rerun the simulations at three different particle density thresholds, $n$ = 10~cm$^{-3}$, $n$ = 1.0~cm$^{-3}$, and $n$ = 0.1~cm$^{-3}$, maintaining an ``early" stellar feedback \citep[stellar winds from massive stars, ][]{stinson13} efficiency, $\epsilon_{\rm ESF} = 13\%$ for the cases of $n$ = 10~cm$^{-3}$ and $n$ = 1.0~cm$^{-3}$, while an efficiency of $\epsilon_{\rm ESF} = 4\%$ is adopted for the lowest density threshold case (Dutton et al. in prep). One galaxy from our sample is shown in Figure \ref{galaxy} at the three different thresholds $n$ explored in this work.

We work towards a quantifiable measure of morphology of the HI gas, which can act as a tracer of stellar feedback activity, testing them on our sample. With a reliable quantity measuring the structure and morphology of galaxies, and an understanding of how this quantity changes with different parameterizations of star formation, we will have developed a tool that can be used to compare simulated galaxies with observational data and later utilized to tune simulations to better agreement with reality.



\section{Quantifying morphology}

\subsection{the Gini coefficient}

The Gini coefficient (or $G$ for short) is a statistical measure of distribution, most commonly used as an economic measure of inequality in a community, where $G=1$ implies maximal inequality and a monopoly on wealth. \citet{lotz04} proposed the adoption of $G$ as a quantifiable measure of galaxy morphology, concluding that galaxies at redshift $z \approx 2$ had higher values of $G$ in comparison to galaxies at $z \approx 0$. \citet{lisker08} studied the effects of signal-to-noise on $G$, and found that $G$ can be highly dependent on the signal-to-noise ratio if signal-to-noise is relatively low. Although this might introduce complications in comparison with observations, this finding poses no problems when it comes to simulated galaxies where background noise doesn't exist. We compute $G$ for our sample of galaxies to explore whether $G$ changes noticeably with a change in star formation threshold but we find no clear correlation.


\subsection{HI bubbles}

Supernova explosions clear out large volumes of space, resulting in what radio astronomers call ``HI bubbles," or ellipsoidal regions of low HI density. A single supernova, however, is highly unlikely to clear out enough gas to result in a resolvable bubble in either observation or simulation. Instead the HI bubbles we observe are believed to be the result of multiple supernovae happening within a short timeframe and in close vicinity with one another. These bubbles, particularly their abundance and size, can serve as a natural proxy for stellar feedback, since they directly (albeit requiring some detective work) correspond to recent supernovae. Ignoring the complication of translating bubble size to the number of supernovae (and stellar winds, in terms of early feedback) and thus to the actual rate of stellar feedback, we develop a simple algorithm to identify density minima in face-on HI maps of our sample of galaxies and grow roughly circular bubbles until the maximum allowed gas density is reached. The threshold we impose is physically arbitrary and motivated by visual inspection of the resulting maps. 

\begin{figure}[h!]
\begin{minipage}{0.5\textwidth}
\includegraphics[width=0.93\textwidth]{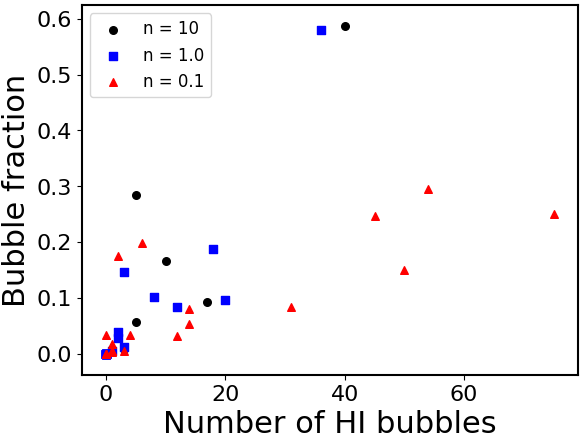}
\caption{\label{fracvholes} The fraction of a galaxy covered by HI bubbles plotted against the number of HI bubbles in a galaxy. Density thresholds, $n$, are in units of cm$^{-3}$. }
\end{minipage}\hspace{1.5pc}%
\begin{minipage}{0.5\textwidth}
\includegraphics[width=0.93\textwidth]{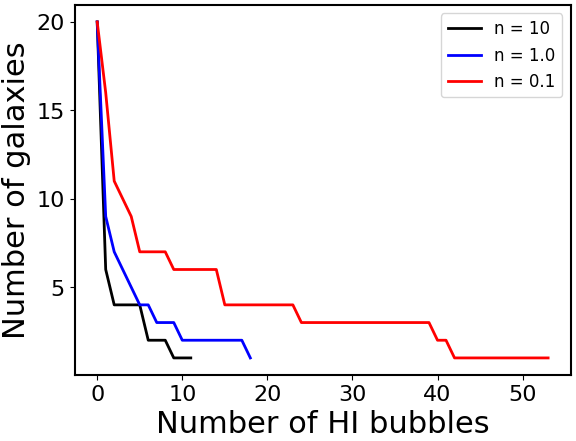}
\caption{\label{holes} We plot the number of galaxies as a function of the minimum number of HI bubbles they contain. Density thresholds, $n$, are in units of cm$^{-3}$.}
\end{minipage}
\end{figure}

We then investigate the number of bubbles in a galaxy and the fraction of the galaxy covered in HI bubbles, restricting our search to the region enclosing $\sim 0.8 M_{\rm gas}$ to ensure that the bubbles lie safely within the galaxy and far from the edge. We find that, as we might intuitively expect, the lowest gas density threshold, $n$ = 0.1~cm$^{-3}$ case produces more small bubbles compared to the higher threshold cases, as shown in Figure \ref{fracvholes}. In Figure \ref{holes} we plot the number of galaxies as a function of the number of galaxies with a minimum number of holes. The curves are distinct enough that a similar search for HI bubbles in observational data could be used to produce similar curves and discriminate between different models of star formation used in cosmological simulations.

\subsection{Radial studies}

\begin{figure}[h]
\begin{center}
\includegraphics[width=0.6\textwidth]{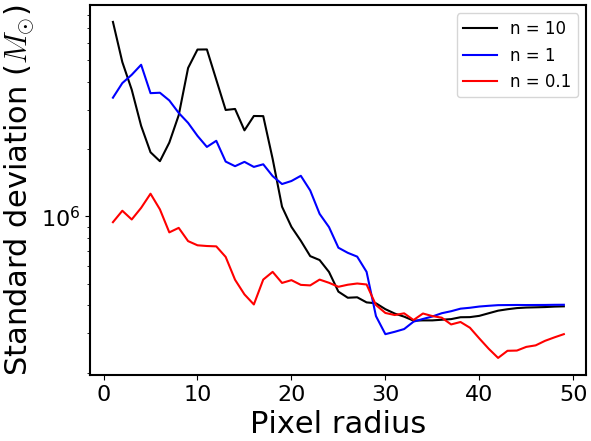}
\end{center}
\caption{\label{anulus} Standard deviation as a function of radius, for the same galaxy shown in Figure \ref{galaxy}}
\end{figure}

We explore the statistical properties of gas density as a function of galaxy radius. Since we use 2D grids to produce maps of galaxies, we use the midpoint circle algorithm to approximate an annulus with the width of one pixel, which we use to compute properties at each `radius'. Radii here are measured in pixels rather than physical units. We plot the standard deviation as a function of pixel radius for each galaxy, finding that $\sigma$ tends to start higher and drop faster for the case of $n$ = 10~cm$^{-3}$, compared to the two other cases. The galaxies with gas density threshold $n$ = 0.1~cm$^{-3}$ start with lower $\sigma$ at the center of the galaxy, but with a slower decrease in standard deviation. One example is Figure \ref{anulus}, of the galaxy shown above in Figure \ref{galaxy}. 
This may simply be a reflection of the observation that galaxies with a lower value of $n$ tend to spread out over a larger radius, as is the case for this galaxy and several others.

\section{Conclusions}
This is project is a work in progress. We hope to soon have a quantity that both encodes useful information about the morphology of a given galaxy and is easy to measure by observational astronomers. Of the three approaches we have tried so far, we find the last method of exploring the properties of a galaxy as a function of radius to be particularly promising. We hope to apply our work to HI observations of real galaxies in the near future.

\subsection{Acknowledgments}
Simulations have been performed on the High Performance Computing resources at New York University Abu Dhabi.
The NIHAO project is a collaboration between NYUAD, the Max Planck Institute for Astronomy in Heidelberg and the
Purple Mountain Observatory in Nanjing.

\newcommand{\newblock}{}

\end{document}